\documentclass[aps,prl,floats,twocolumn,showpacs,superscriptaddress]{revtex4}

\usepackage{graphicx}
\usepackage{rotate}
\begin{document}
\topmargin 0.5cm.
\title{Synchronization reveals topological scales in complex networks}

\author{Alex Arenas}

\affiliation{Departament d'Enginyeria Inform{\`a}tica i Matem{\`a}tiques,
  Universitat Rovira i Virgili, 43007 Tarragona, Spain}

\author{Albert D{\'\i}az-Guilera}

\affiliation{Departament de F{\'\i}sica Fonamental, Universitat de
  Barcelona, Marti i Franques 1, 08028 Barcelona, Spain}

\author{Conrad J. P{\'e}rez-Vicente}

\affiliation{Departament de F{\'\i}sica Fonamental, Universitat de
  Barcelona, Marti i Franques 1, 08028 Barcelona, Spain}

\date{\today}

\begin{abstract}

We study the relationship between topological scales and dynamic time scales in complex networks. The analysis is based on the full dynamics towards synchronization of a system of coupled oscillators.
In the synchronization process, modular structures corresponding to well defined communities of nodes emerge in different time scales, ordered in a hierarchical way. The analysis also provides a useful connection between synchronization dynamics, complex networks topology and spectral graph analysis.

 \end{abstract}

\pacs{xxx.xxx}

\maketitle

The science of complex networks has been a subject of attention of
the physicists' community in the recent years
\cite{strogatz, barev, yamirrep}. Complex
networks are found in fields as diverse as the Internet, the
World-Wide-Web, food-webs, and biological and social organizations
(see \cite{buchanan} and references therein). Although the main
characteristics of complex networks have been properly described
at the microscale level (node properties) and also at the macroscale level
(whole network properties) some of the characteristics of the mesoscale
are still elusive. In particular, the community detection problem
concerning the determination of mesoscopic structures that have
functional, relational or even social entity is still
controversial, starting  from the 'a priori' definition of what a
community is \cite{newmanepjb,jstat}.

The community detection problem consists in finding a 'good'
partition of the network in sub-graphs that represent communities
according to a given definition. However, in many complex networks
the organization of nodes is not completely represented by a
unique partition but by a set of nested communities that appear at
different topological scales. Let us consider as a naive example
the network formed by all human acquaintances. Thus, at some
topological scale we can expect to find many communities formed by
families, friends and soon, beyond this scale the expected
partitions into cities will come up, beyond this regions, after
that countries, and finally probably continental areas. Here, we are aimed at giving a
method to reveal these different topological scales.

In a completely different scenario, physicists have largely
studied the dynamics of complex biological systems, and in particular the
paradigmatic analysis of large populations of coupled oscillators \cite{winfree,strogatzsync, kurabook}.
The emergence of synchronization patterns in these systems has
been shown to be closely related to the underlying topology of
interactions. In this letter we show that, for a suitable model,
the dynamical process towards synchronization shows different
patterns over time intrinsically connected with the hierarchical
organization of communities in complex networks. The ubiquity of
synchronization phenomena in real world makes appealing this
approach from a physical and biological perspective. Moreover we
will show that the connections with the spectral theory of the
Laplacian matrix of a graph spreads the possibilities of the
analysis to any complex network.

One of the most successful attempts to understand synchronization
phenomena was due to Kuramoto \cite{kurabook}, who analyzed a model
of phase oscillators coupled through the sine of their phase
differences. The model is rich enough to display a large variety
of synchronization patterns and sufficiently flexible to be
adapted to many different contexts \cite{conradrev}. The Kuramoto model consists of a population of $N$ coupled phase oscillators where the phase of the $i$-th unit, denoted by $\theta_i(t)$, evolves
in time according to the following dynamics
\begin{equation}
\frac{d\theta_i}{dt}=\omega_i + \sum_{j}
K_{ij}\sin(\theta_j-\theta_i) \hspace{0.5cm} i=1,...,N
 \label{ks}
\end{equation}
\noindent where $\omega_i$ stands for its natural frequency and $K_{ij}$ describes the coupling between units. The
original model studied by Kuramoto assumed mean-field interactions
$K_{ij}=K, \forall i,j$. If the oscillators are identical $(\omega_i = \omega ~\forall i)$ there is only one attractor of the dynamics: the fully synchronized regime where $\theta_i = \theta, ~\forall i$. 
Recently, due to the realization that many networks
in nature have complex topologies, these studies have been extended 
to complex networks with local interaction \cite{barahona,motter1,yamir,hong,motter2,lee,chavez,munozprl}.

In particular, it has been shown \cite{yamir2,kahng} that high densely interconnected sets of oscillators (motifs) synchronize more easily that those with sparse connections.
This scenario suggests that for a complex network with a non-trivial
connectivity pattern, starting from random initial conditions,
those highly interconnected units forming local clusters will
synchronize first and then, in a sequential process, larger and
larger spatial structures also will do it up to the final state where the whole population should have
the same phase. We expect this process to occur at different time scales if a clear community structure exists.  Thus, the dynamical route towards the global
attractor will reveal different topological structures, presumably those which represent communities.
Therefore, it is the complete dynamical process what unveils the
whole organization at all scales, from the microscale at a very
early stages up to the macroscale at the end of the time evolution. On the contrary, those systems endowed with a regular topological structure will display a trivial dynamics with a
single time scale for synchronization. 

To study this phenomena, instead of considering a global observable, we
define a local order parameter measuring the average of the
correlation between pairs of oscillators
 \begin{equation}
\rho_{ij}(t)=<cos(\theta_i(t)-\theta_j(t))>
 \label{ro}
\end{equation}
\noindent where the brackets stand for the average over initial random phases.
The main advantage of this approach is that it allows to trace the time evolution of pairs of
oscillators and therefore to identify compact clusters reminiscent
of the existence of communities. 

To give evidence of the
aforementioned facts we have analyzed the dynamics towards
synchronization --time evolution of $\rho_{ij}(t)$-- in computer-generated graphs
with a hierarchical community structure. 
In \cite{girvannewman} the authors proposed models of networks
with a well defined community structure, that have been used as a benchmark 
for different community detection algorithms \cite{jstat}.
Here, we propose a generalization of this model that includes two hierarchical levels 
of communities. The graphs we generate are as follows: we prescribe, in a set of 256 nodes, 16 compartments that will represent our first community organizational level, and four compartments containing each one four different compartments of the above first level, that define the second organizational level of the network.
The internal degree of nodes at first level $z_{in_1}$ and the internal degree
of nodes at second level $z_{in_2}$ keep an average degree
$z_{in_1}+z_{in_2}+z_{out}=18$. From now on, networks with two
hierarchical levels are indicated as $z_{in_1}$ - $z_{in_2}$, e.g.
a network with 13-4 means 13 links with the nodes of its first
hierarchical level community (more internal), 4 links with the
rest of communities that form the second hierarchical level (more
external) and 1 link with any community of the rest of the
network.

\begin{figure}
 \includegraphics*[width=0.47\columnwidth]{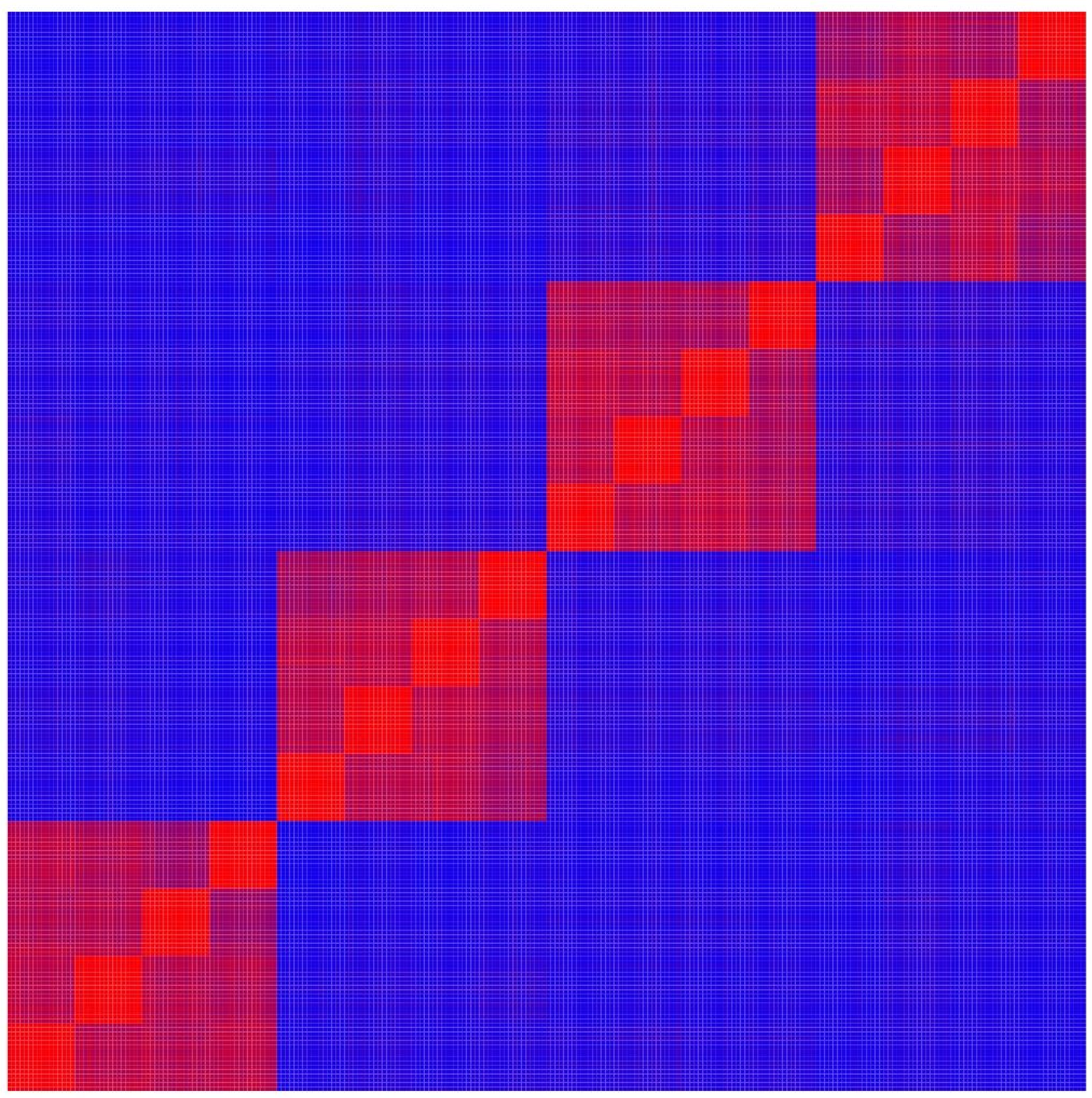} \hfill
  \includegraphics*[width=0.47\columnwidth]{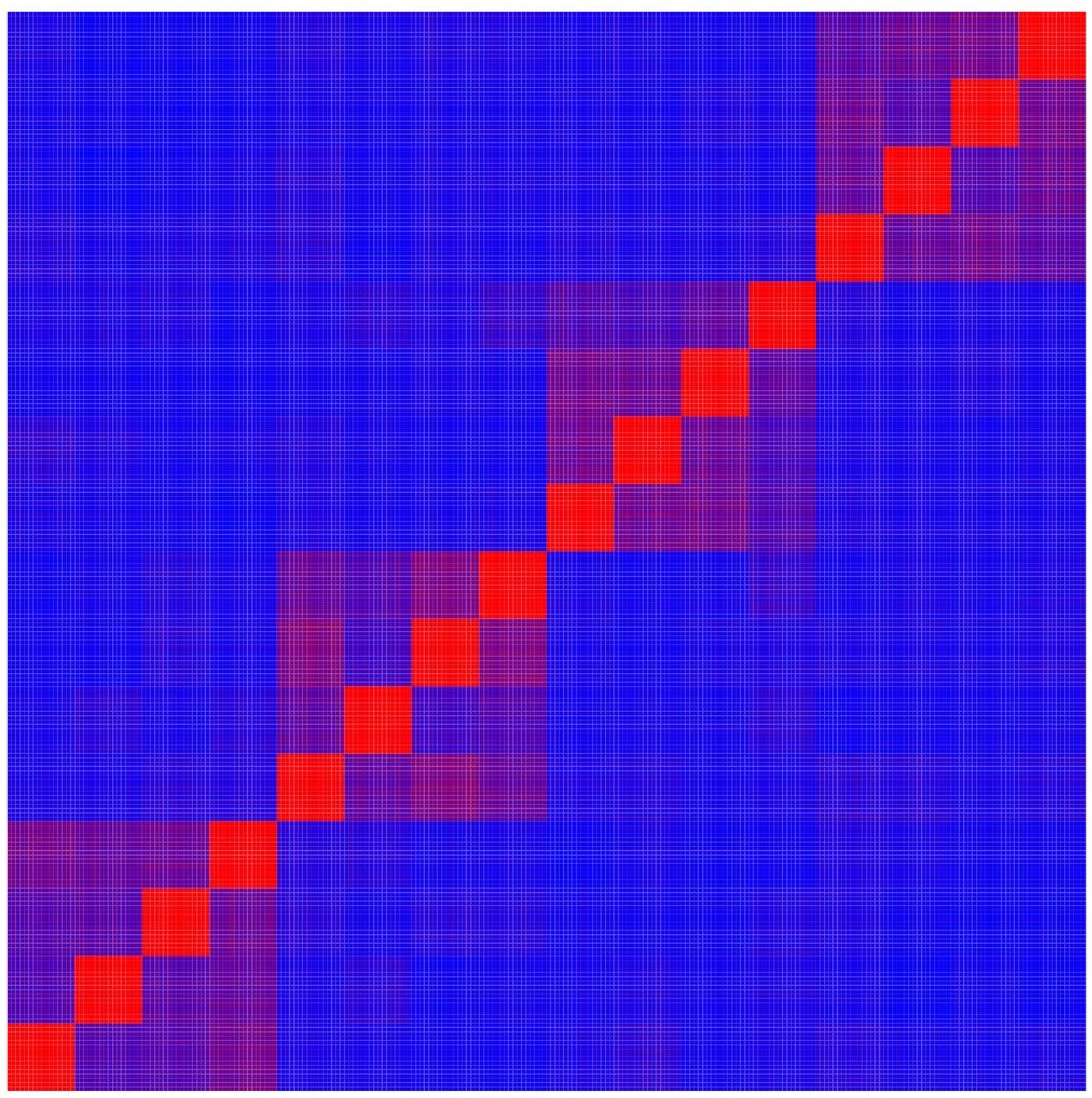} 
  \caption{Color on-line. Average of the correlation between pairs of oscillators.
  The structure networks are 13-4 (left) and 15-2 (right). See text for a description of 
  the networks. The colors are a gradation between blue (0) and red (1).}
\label{corr_mat}
\end{figure}

In Fig. \ref{corr_mat} we represent $\rho_{ij} (t)$ at the same time $t$
for two slightly different hierarchical networks 13-4 and 15-2.   
In the two figures we can identify the two levels of the hierarchical distribution of communities.
The network 13-4 (left) is very close to a state in which the four large groups are almost synchronized
whereas the network 15-2 (right) still presents some of the smaller groups of 
synchronized oscillators, and the larger group starting to synchronize, coherently with their topological structure.

The visualization of the correlation matrix of the system
helps in elucidating the topology of the network. To extract the quantitative information it is useful 
to introduce some threshold $T$ to convert the correlation matrix into a binary matrix, that will be used to determine the borders between different groups. We define a {\em dynamic  connectivity} 
matrix  
 \begin{equation}
{\cal D}_t (T)_{ij}=
\left\{
\begin{array}{ll}
1 & \mbox{if } \rho_{ij}(t)>T \\
0 & \mbox{if } \rho_{ij}(t)<T
\end{array}
\right. 
 \label{ro_threshold}
\end{equation}
that depends on both the underlying topology and the collective dynamics. For a fixed time $t$, by moving the threshold $T$, we obtain different representations of ${\cal D}_t(T)$ that inform about the structure of the dynamic correlations. When the threshold is large enough the representation of ${\cal D}_t(T)$ becomes a set of disconnected clumps or communities. Decreasing $T$ a hierarchical structure of communities is devised. Note that 
since the function $\rho_{ij}(t)$ is continuous and monotonic (because the existence of a unique attractor of the dynamics), we can redefine ${\cal D}_T(t)$, i.e. fixing the threshold and evolving in time. We obtain the same information about the structure of the dynamic connectivity matrix at different time scales. Let us show that these time scales unravel the topological structure of the connectivity matrix at different topological scales.

\begin{figure}
 \includegraphics*[angle=270,width=\columnwidth]{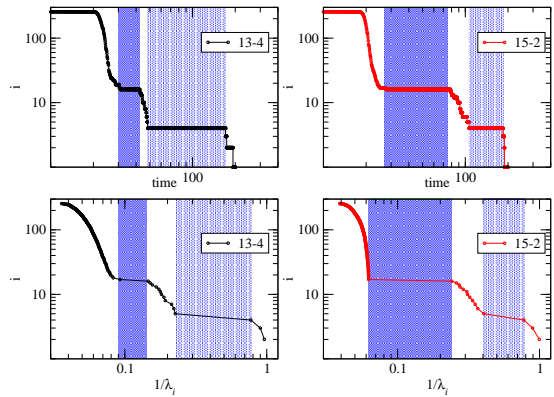} 
  \caption{Color on-line. Top: Number of connected synchronized components (equivalent to number of null eigenvalues of $S({\cal D}_T(t))$) as a function of time 
  for the two networks of Fig. \ref{corr_mat} at $T=0.99$. Bottom: Rank index $i$ (see text) versus the inverse of the corresponding eigenvalues of the Laplacian matrix  ${\cal L}$. The shadow regions indicate the stability plateaus for 16 (dark) and 4 (light) communities. The same representation is used for the plateaus in the eigenvalue spectrum corresponding to indices 16 and 4.}
\label{eigenvalues}
\end{figure}

From the eigenvalue spectrum of ${\cal D}_T(t)$, $S({\cal D}_T(t))$, one can extract the number of disconnected components of the system as the number of null eigenvalues. The evolution of $S({\cal} D_T(t))$ traces the hierarchy of communities as follows: at short times, all units are uncorrelated and then we have $N$ disconnected sets, being $N$ the number of nodes in the network; as time goes on, nodes become synchronized in groups according to their topological structure. In Fig. \ref{eigenvalues} (top) we plot, for the two networks analyzed in  Fig. \ref{corr_mat}, the number of disconnected components as a function of time, for a fixed threshold $T$. We can observe the relative stability of the
two partitions for the two networks, corresponding to the two prescribed hierarchical levels. For the 13-4 network the synchronization of the 4 groups of 64 nodes each is much more stable than the 16 groups of 16 nodes, i.e. the community structure at the second hierarchical level is stronger, whereas the opposite can be inferred for network 15-2. 

Another interesting link between dynamics and topology can be highlighted from the analysis of the whole spectrum of the Laplacian matrix of the network graph ${\cal L}$ \cite{biggs}. The spectral information of the Laplacian matrix has been used to understand the structure of complex networks \cite{barabasispect}, and in particular to detect the community structure \cite{donetti,capocci}. Recent studies have also focused on the spectral information of the Laplacian matrix and the synchronization dynamics \cite{barahona,motter1,yamir,hong,motter2,lee,chavez, munozprl}. The common approach is to take advantage of the master stability equation \cite{pecora} to determine the relation between the relative stability of the synchronized state (via the ratio $\lambda_N/\lambda_2$) and the heterogeneity of the topology, although sometimes some language abuse appears and authors talk about better or worse synchonizability instead of stability of the synchronized state. Our approach differs from these works in the following: we are interested in the transient towards synchronization because it is this whole process which will reveal the topological structure at different scales. For this reason our analysis focus on the whole eigenvalue spectrum of the Laplacian matrix $S({\cal L})$. 

To characterize this spectrum, we rank the eigenvalues of ${\cal L}$ using an index $i$ in ascending order $0=\lambda_1\le  \lambda_2\le \ldots \lambda_i \ldots \le \lambda_N$. The structure of this sequence brings to light many aspects of the topological structure:  (i) the number of null eigenvalues gives trivially the number of disconnected components, (ii) the gaps between consecutive eigenvalues tell us about the relative differences of time scales, and (iii) large eigenvalues in the last part of the series stands for the existence of hubs in the network (we will turn to these points later).  In Fig. \ref{eigenvalues} (bottom) we have plotted the eigenvalues of the Laplacian matrix for the 13-4 and 15-2 structures. We observe three groups of eigenvalues separated by gaps. Each gap separates a community either of 256 groups, 16 groups, 4 groups elements or the whole population. 
Notice that for the
13-4 graph the plateau of 16 communities is shorter than the plateau for 4 communities and the contrary  for the 15-2 case, indicating that the 16 clusters community is less well defined in the former case. Indeed, the ratio between the eigenvalues  
is a good quantitative measure of the stability of the structure (which is measured in terms of modularity in other studies \cite{jstat}) and is related to the length of the plateaus observed in Fig. \ref{eigenvalues} (top).

\begin{figure}
\includegraphics*[height=50mm, width=0.4\columnwidth]{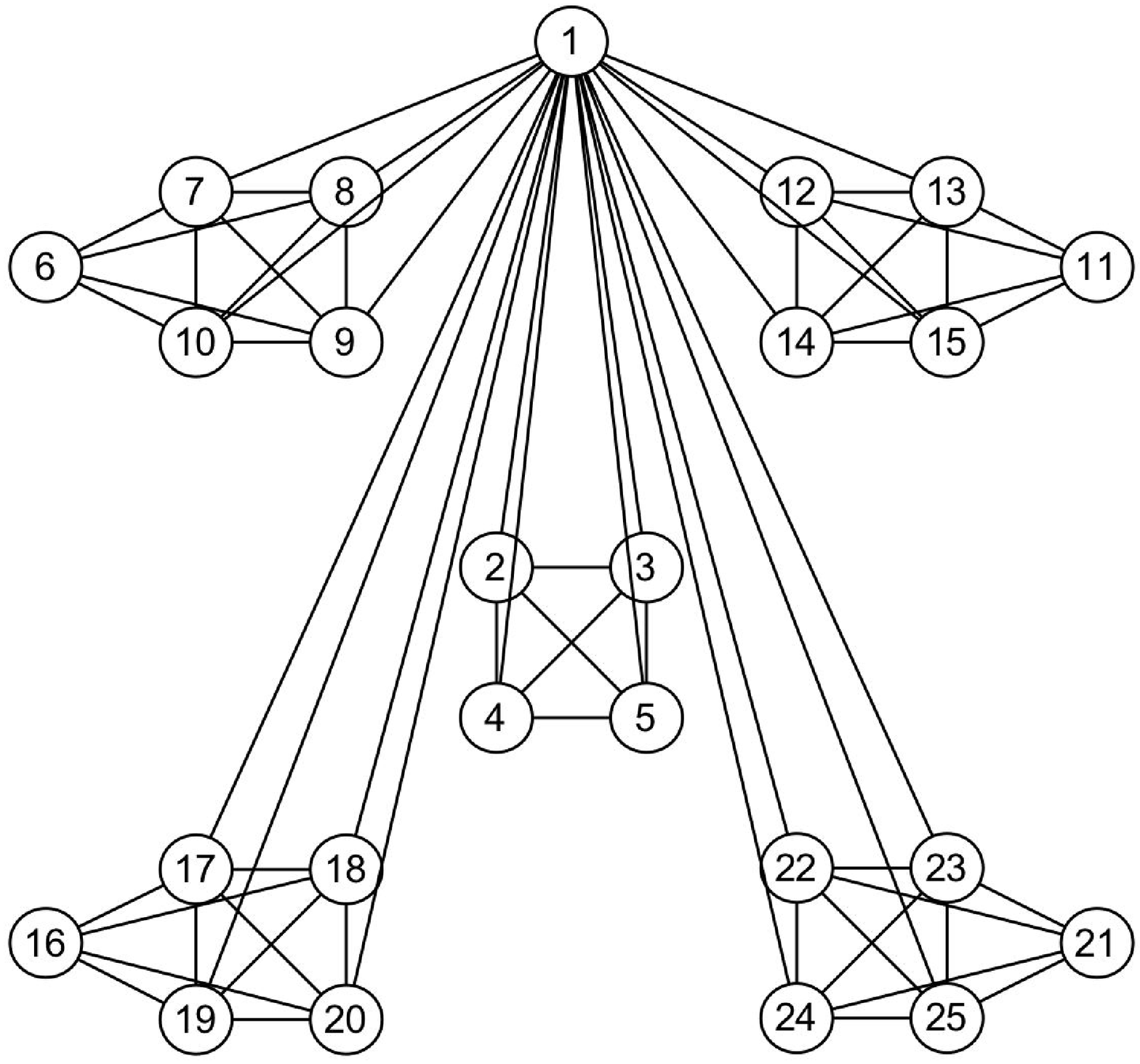} \hfill
 \includegraphics*[width=0.55\columnwidth]{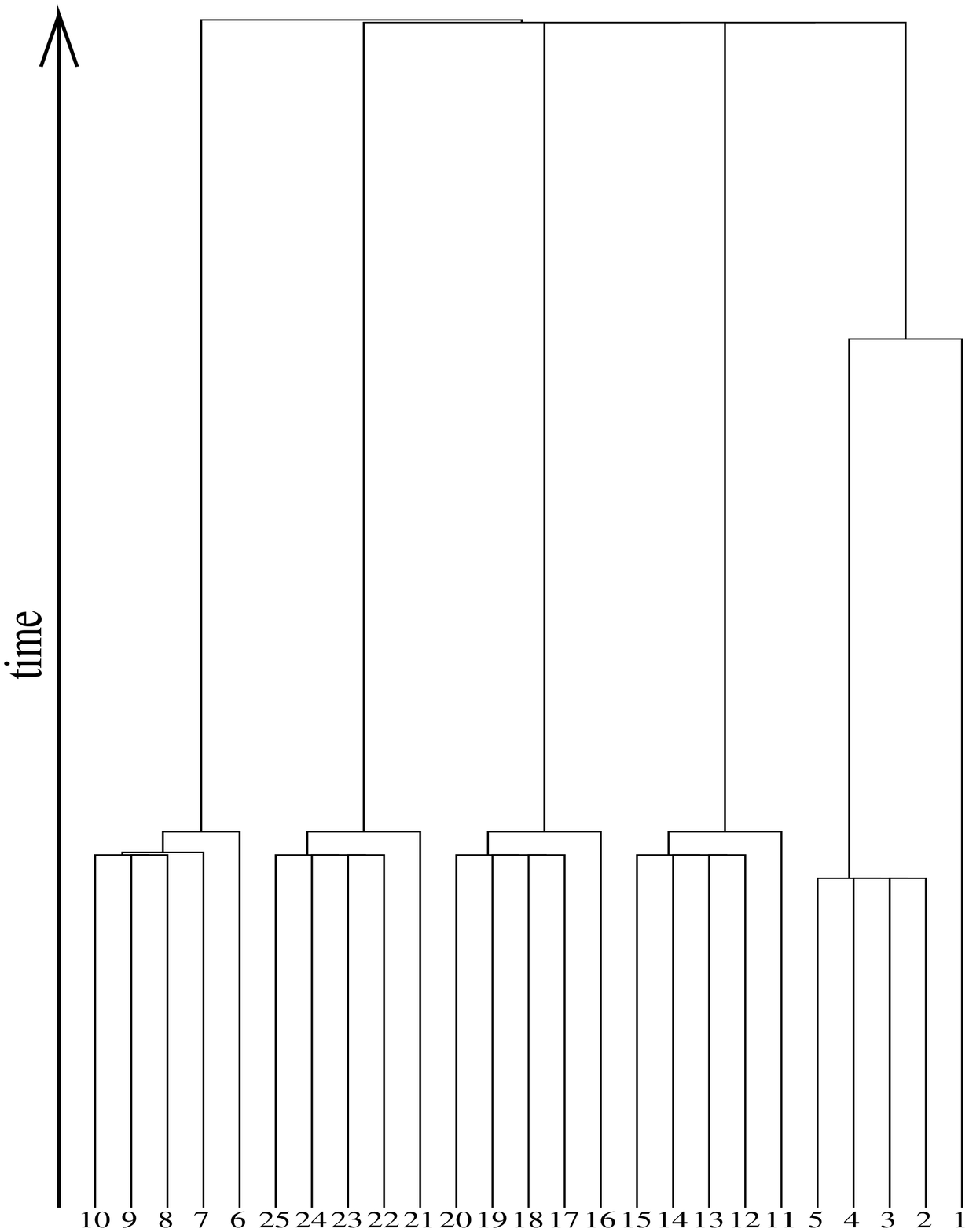} 
  \caption{Left: RB network of 25 labeled nodes with two hierarchical levels.
  Right: Time evolution of the synchronization process between labeled oscillators. The length of the dendogram branches indicate the relative stability of the different structures.}
\label{dendo}
\end{figure}

We visualize the formation of the connected groups of synchronized oscillators
in time constructing a dendogram in which we draw lines between groups of oscillators when they merge. Applying this technique to the above defined networks we can see two different topological scales disclosed by synchronization and the relative stability of them. The networks investigated so far are homogeneous in degree. At this point we ask about the effect when inhomogeneities in degree are considered. We have applied this procedure to the network structure proposed by Ravasz and Barabasi \cite{RB} with a hierarchical structure in two levels and a scale-free degree distribution. As can be seen from the dendogram depicted in Fig. \ref{dendo} the communities synchronize at different times, depending on its role in the hierarchy, and it also shows the remarkably effect of hubs in the synchronization process.

Finally we would like to shed some light about the intriguing relationship between the  eigenvalues of the Laplacian and the dynamic structures that emerge towards synchronization. To understand this correspondence let us analyze the linearized dynamics of the Kuramoto model (i.e. the dynamics close to the attractor of synchronization) in terms of the Laplacian matrix, 
\begin{equation}
\frac{d\theta_i}{dt}=-k\sum_{j}
L_{ij} \theta_j \hspace{0.5cm} i=1,...,N
 \label{linearmodel}
\end{equation}
whose solution in terms of the normal modes $\varphi_i(t)$ reads
\begin{equation}
\varphi_i(t)=\sum_{j} B_{ij}\theta_j= 
\varphi_i(0) e^{-\lambda_i t}\hspace{0.5cm} i=1,...,N
 \label{linearsolution}
\end{equation}
where$\lambda_i$ are the eigenvalues of the
Laplacian matrix, and $B$ is the eigenvectors matrix. 

This set of equations has to be satisfied at any time $t$. If we rank the system of equations in descending order of the eigenvalues (i.e. starting from $\lambda_N$), the right hand side system of Eq.(\ref{linearsolution}) will approach zero in a hierarchical way. This fact is equivalent in the dynamics to group oscillators surpassing the synchronization threshold forming communities. The gaps in the spectrum $S({\cal L})$ represent clearly different time scales between modes revealing different topological scales. The collective modes, solution of the system represented by Eq.(\ref{linearsolution}), denote two types of behaviors. Some modes provide information about reorganization of the phases in the whole network, while the others inform about synchronization between pairs or groups of oscillators. The presence of hubs in the topology gives rise to large eigenvalues that decay very fast and are related to the first type of modes, those representing "synchronization" between the hub and the {\em topological average} of the phases of rest of oscillators. 
The rest of modes relate oscillators that have similar projections on the corresponding eigenvectors thus giving rise to communities at a given topological scale. Indeed, this fact support the success of the identification of communities using spectral analysis \cite{donetti}.

Summarizing, we have analyzed the synchronization dynamics in complex networks and show how this process unravels its different topological scales. We have also reported a connection between the spectral information of the Laplacian matrix and the hierarchical process of emergence of communities at different time scales.

\begin{acknowledgments}
  We thank  M. A. Mu\~{n}oz, Y. Moreno and R. Guimer\`{a} for helpful comments. This work has been supported by DGES of the Spanish Government Grant No. BFM-2003-08258 and EC-FET Open Project No. IST-2001-33555.
\end{acknowledgments}

\end{document}